\documentclass[%
 aip,
 amsmath,amssymb,
 reprint,%
]{revtex4-1}

\usepackage{graphicx}
\usepackage{dcolumn}
\usepackage{bm}

\usepackage[utf8]{inputenc}
\usepackage[T1]{fontenc}
\usepackage{mathptmx}
\usepackage{etoolbox}
\usepackage{xcolor,soul}
\makeatletter
\def\@email#1#2{%
 \endgroup
 \patchcmd{\titleblock@produce}
  {\frontmatter@RRAPformat}
  {\frontmatter@RRAPformat{\produce@RRAP{*#1\href{mailto:#2}{#2}}}\frontmatter@RRAPformat}
  {}{}
}%
\makeatother
\begin{document}

\preprint{AIP/123-QED}

\title{Thawed Matrix method for computing Local Mechanical Properties of Amorphous Solids}

\author{J\"org Rottler$^*$}
\email{jrottler@physics.ubc.ca}
\affiliation{Department of Physics and Astronomy and Quantum Matter Institute, University of British Columbia, Vancouver, British Columbia, V6T 1Z1, Canada }%

\author{C\'eline Ruscher}%
\affiliation{ Department of Mechanical Engineering, University of British Columbia, Vancouver, British Columbia, V6T 1Z4, Canada}%

\author{Peter Sollich} 
\affiliation{
Institute for Theoretical Physics, Georg-August-Universit\"at G\"ottingen, 37077 G\"ottingen, Germany}%

\date{\today}

\begin{abstract}
We present a method for computing locally varying nonlinear mechanical properties in particle simulations of amorphous solids. Plastic rearrangements outside a probed region are suppressed by introducing an external field that directly penalizes large nonaffine displacements. With increasing strength of the field, plastic deformation can be localized. We characterize the distribution of local plastic yield stresses (residual local stresses to instability) with our approach, and assess the correlation of their spatial maps with plastic activity in a model two-dimensional amorphous solid. Our approach reduces artefacts inherent in a previous method known as the 'frozen matrix' approach that enforces fully affine deformation, and improves the prediction of plastic rearrangements from structural information.
\end{abstract}

\maketitle

\section{Introduction}
Despite being macroscopically isotropic, amorphous solids exhibit structural heterogeneities at the nanoscale \cite{yoshimoto2004mechanical, tsamados2009local,widmer2008irreversible,tanguy2010vibrational,manning2011vibrational,zylberg2017local, schwartzman2019anisotropic}. Accurate methods for assessing local, spatially varying elastic and plastic properties are important for linking particle and coarse grained descriptions \cite{rodney2011modeling,nicolas2018deformation,puosi2015probing,castellanos2021insights} of such materials. Distributions of local elastic tensors can be computed nonperturbatively using fluctuation (linear response) formulas, which requires a choice of coarse-graining length scale. While the mean of such distributions is always close to the macroscopic value, the relative fluctuations about the mean decrease inversely with the coarse-graining scale (in two dimensions) \cite{tsamados2009local}.  In model Lennard-Jones glasses, a characteristic length scale of five particle diameters was identified below which linear elasticity is no longer valid. 

The computation of nonlinear plastic properties, i.e.~the distance of a local region from mechanical instability, is more complicated. In particle based simulations, it requires deforming the solid but suppressing plastic activity everywhere in the material except in the local region that is to be characterized. This task requires some form of constraint on the surrounding material. One possibility is the so-called ``frozen matrix'' approach, in which the particles in the surrounding material are restricted to undergo purely affine deformation, thus precluding irreversible (i.e.~locally non-affine) rearrangements in the matrix. This method was proposed by one of us \cite{frozen} and has since been deployed in a variety of contexts to assess local yield stress distributions \cite{puosi2015probing,shang2018role}, their evolution under shear deformation \cite{ruscher2020residual} and the correlations between such local yield stresses and plastic activity \cite{patinet2016connecting,barbot2018local,richard2020predicting}. While these studies have focused mostly on athermal quasistatic deformation of two dimensional packings, applications to three-dimensional glasses that have explored the dynamical moduli of supercooled fluids \cite{shang2019local} or a three-dimensional yield surface have also been performed \cite{ruan2022predicting}.

Notwithstanding the well-documented success of the frozen matrix method, it was pointed out early on that the truncation of the nonaffine displacement field outside the probed region introduces rather severe artefacts that become visible already at the linear response level \cite{mizuno2013measuring}. In general, larger shear moduli are observed because no relaxation is possible in the frozen region. Similarly, one can expect that local yield stresses are being overestimated and that their distributions become artificially compressed \cite{ruscher2020residual}.  For this reason, there is currently renewed interest in developing avenues for improvements. For instance, the recently introduced ``soft matrix'' approach proposes to couple particles in the matrix to an affinely moving background using linear springs, thus allowing some degree of nonaffinity to occur \cite{adhikari2023soft}. This methodology was used to extract a plasticity length scale from the dependence of the mean yield strain on the linear dimension of the local region. While this method without doubt is suitable for interpolating between an unconstrained matrix and a fully frozen one, the extent of suppression of plastic activity or the correlation with local plastic activity, both of which must depend on the strength of the effective constraint from the springs, has not yet been studied systematically.

In this contribution, we explore an alternative possibility to improve over the frozen matrix method that we term 'thawed matrix'. Similar to Ref.~\onlinecite{adhikari2023soft}, we aim to alleviate the frozen matrix artefacts by allowing a degree of nonaffine particle motion in the surrounding matrix. Instead of using a harmonic tethering to an affine reference configuration, we target more specifically large localized nonaffine displacements that are known to occur at the core of plastic rearrangements. Such a strategy has previously been proposed as a means of controlling defect formation in colloidal crystals \cite{ganguly2013nonaffine,ganguly2015statistics,nath2018existence}. It targets directly the suppression of strong {\em local} nonaffinity in the matrix, rather than penalizing large displacements from an affine reference \cite{adhikari2023soft} that are known to occur in {\em large-scale} nonaffine deformation patterns around plastic rearrangements, forming vortex-like patterns with correlations extending over many particle diameters \cite{leonforte2005continuum}.

In what follows, we describe and present a thorough characterization of the performance of our proposal for computing elastic moduli and local yield stresses, their distributions and spatial maps. We also benchmark the predictions of local plastic activity from our 'thawed matrix' against the traditional 'frozen matrix' approach. Although our approach is not able to fully remove all constraint artefacts, it provides local yield stress maps that capture more faithfully the ensuing plastic activity.

\section{Thawed Matrix Method}

The squared nonaffine residual displacement $D_{min}^2(k)$ of a particle $k$ between two configurations is an excellent indicator of local plastic activity \cite{falk1998dynamics}.
This 'nonaffinity parameter' \cite{ganguly2013nonaffine} is given by the sum of squares of all the residual displacements
of the particles in some local region relative to the best possible affine deformation $\epsilon_{\alpha\beta}$,
measured with respect to a reference configuration. Denoting by $d_{nk}^\alpha=r_n^\alpha-r_k^\alpha$ and $D_{nk}^\alpha=R_n^\alpha-R_k^\alpha$ the distance between a particle $n$ in the neighborhood of a central particle $k$ in the actual (lower case) and reference (upper case) configurations, one computes
\begin{equation}    
D_{min}^2(k)=\sum_n\sum_\alpha\left[d_{nk}^\alpha-\sum_\beta(\epsilon_{\alpha\beta}+\delta_{\alpha\beta})D_{nk}^\beta\right]^2
\label{d2min-eq}
\end{equation}
with $\epsilon_{\alpha\beta}=\sum_\gamma X_{\alpha\gamma}Y_{\beta\gamma}^{-1}-\delta_{\alpha\beta}$, $X_{\alpha\beta}(k)=\sum_nd_{nk}^\alpha\cdot D_{nk}^\beta$ and $Y_{\alpha\beta}(k)=\sum_nD_{nk}^\alpha\cdot D_{nk}^\beta$. The the sum is taken over all particles in the neighborhood, whose radius we set equal to the truncation range $r_c$ of the interatomic potential. In order to locally suppress plastic activity, we propose to add an additional energy penalty
\begin{equation}
    U_{NA}=h\sum_k^ND_{min}^2(k)
    \label{tm-eq}
\end{equation} 
where the sum runs over those particles that are not supposed to experience plastic events. Clearly $h\rightarrow \infty$ enforces purely affine deformation, while $h=0$ yields unconstrained motion.

\begin{figure}[t]
\includegraphics[width=0.8\columnwidth]{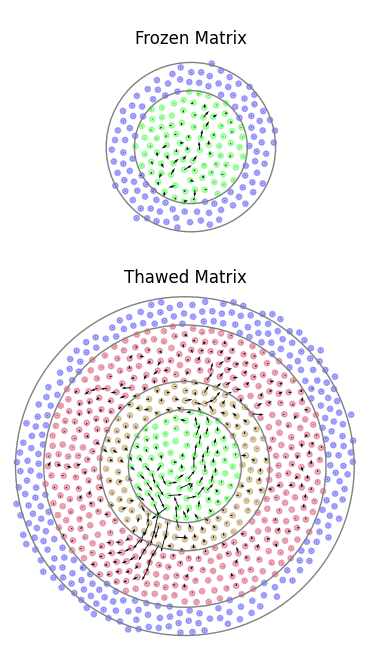}
\caption{\label{fig:illu} Illustration of the frozen and thawed matrix methods. Particles in an inner (green) region for $r<5$ are completely unconstrained, while particles in an outer (blue) region are frozen and move only affinely. In the frozen matrix method, the nonaffine displacements (black arrows) are immediately truncated at $r=5$, while in the thawed matrix approach, a buffer region (red and brown) for $5\le r \le 12.5$ is inserted, in which particles are unconstrained but feel an additional force from $U_{NA}$ that suppresses localized nonaffine motion. Both transition (brown) and frozen (blue) regions have a width equal to $r_c=2.5$.}
\end{figure}

Ideally, to probe the nonlinear mechanical properties of some local region, one would simply apply this additional potential to all but those particles in the region of interest. The computational effort, however, grows proportionally to the number of particles in the simulation, and there is an additional overhead coming from computing forces resulting from eq.~\eqref{tm-eq}. As a compromise, we propose the 'thawed matrix' method illustrated in Fig.~\ref{fig:illu} as a potential improvement over the 'frozen matrix' approach. In the latter, only particles in a central circular region of interest (green) plus a frozen region (blue) are kept in the simulation. By construction, this method suppresses entirely all nonaffine displacements outside the region being probed. To reduce the artefacts of this truncation, we introduce a buffer (red and brown) region, in which particles relax freely but feel additional forces coming from eq.~\eqref{tm-eq}. The sum in eq.~\eqref{tm-eq} only includes particles in the red region. However, particles in the brown region still feel some constraint force as they are part of the sum of eq.~\eqref{d2min-eq}. This creates a natural transition region, in which the nonaffine suppression forces ramp up from zero in the green region to full strength in the red region. Importantly, the nonaffine displacement field is thus allowed to extend continuously into the buffer region. A final, outer boundary (blue) is retained as fully frozen to prevent rearrangements at the boundary. 

\section{Computational Details}
For our numerical demonstration, we use a two-dimensional binary 65:35 Lennary-Jones system suggested by Br\"uning et al. \cite{bruning2008glass}. Systems containing 20,000 particles are prepared at fixed density $\rho=1.2$ (all quantities in reduced Lennard-Jones units) and briefly equilibrated for $500\,\tau_{\rm LJ}$ at temperature $T=1$, before being subjected to cooling to zero temperature with cooling rate $\dot{T}=(50,000\,\tau_{\rm LJ})^{-1}$. This yields a (by conventional molecular dynamics standards) reasonably well annealed amorphous packing. Mechanical properties of these quenched states are investigated using the standard athermal quasistatic (AQS) simple shear protocol, which consists of repeatedly straining the system by a small strain increment $\delta \gamma=5\times 10^{-5}$, followed by energy minimization with the conjugate gradient algorithm. The LAMMPS code \cite{thompson2022lammps} is used for all simulations.

To assess the spatial variation of local mechanical properties in a computationally efficient manner, we follow ref. \onlinecite{barbot2018local} and partition the systems on a grid with spacing equal to the potential cutoff $r_c =2.5$. Unless otherwise stated, both local shear moduli and local yield stresses are evaluated in circular regions of radius $r=5$, which was suggested by Patinet et al. as a value that optimizes correlations with plastic activity \cite{patinet2016connecting} and is also a lower bound for the applicability of Hooke's law \cite{tsamados2009local}. Elastic shear moduli for each region $i$ are obtained by reading off the stress increment resulting from straining the regions by a total strain of 0.2\%. We take the first stress drop in the AQS protocol as our definition of plastic yield, and the local yield stress $\Delta \tau_i$ is taken as the total change in local shear stress up to that point relative to the original quenched configuration. Again following Patinet et al.~\cite{patinet2016connecting}, we perform an additional optimization by considering multiple loading directions $\alpha$ and projecting onto the direction of remote loading $(\alpha=0)$, i.e. $\Delta \tau_{y,i}=\rm{min}_\alpha [\Delta \tau_i(\alpha)/\cos(2\alpha)]$. 

\section{Results}
\begin{figure}[t]
\includegraphics[width=0.8\columnwidth]{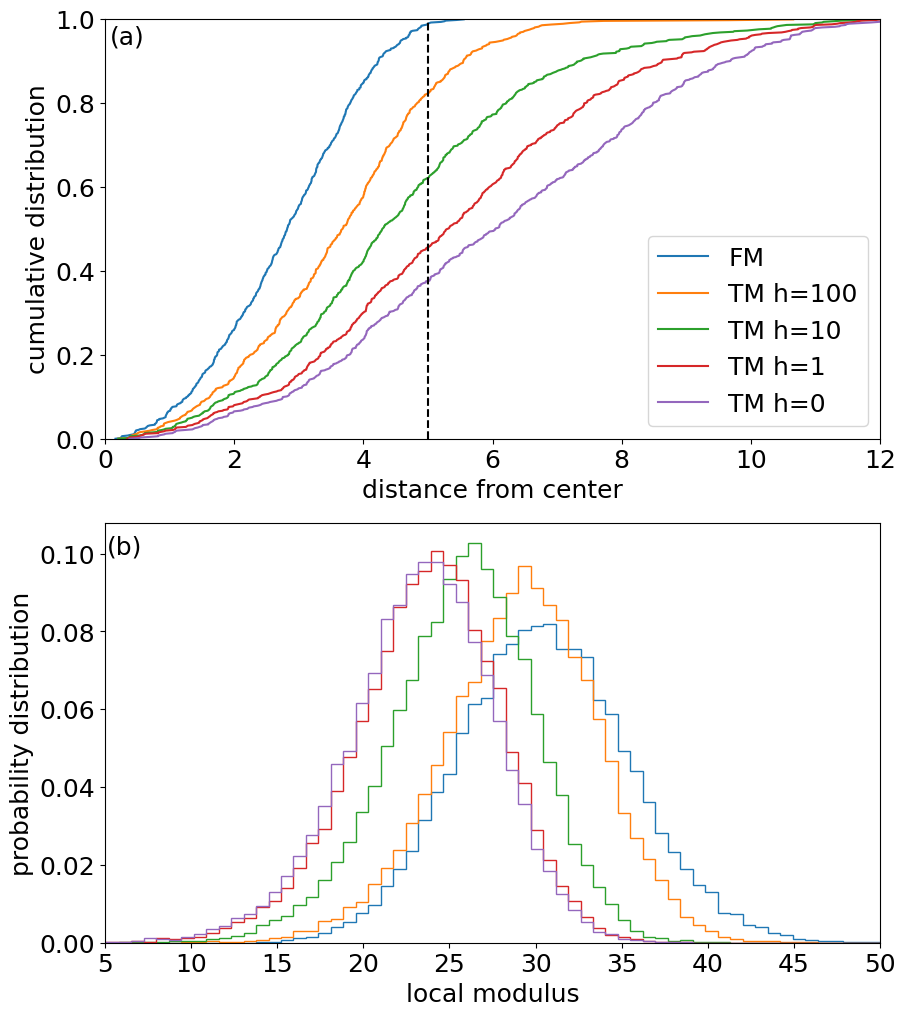}
\caption{\label{fig:distmod} (a) Cumulative probability distribution of the distance of first plastic event to the center of the probed region obtained by the frozen matrix (FM) and thawed matrix (TM) methods. (b) Probability distribution of local (linear) shear moduli.}
\end{figure}

Mapping out the local yield stress distribution of an amorphous material is a task that requires optimization of two conflicting goals: one needs to localize plastic activity in a region as small as possible in order to reveal local variation, while minimally perturbing the system overall. In order to guide the optimal choice of the coupling parameter $h$, we first assess how well the thawed matrix (TM) method suppresses plastic activity for different values of $h$. To this end, we deform approximately 600 regions until a plastic event is detected. We then compute the $D_{min}^2(k)$ observable for all particles $k$ (using a local environment of radius 2.5), and consider its maximum as the center of the plastic shear transformation. Fig.~\ref{fig:distmod}(a) reports the cumulative probability distribution for finding the plastic event a distance $r$ from the center of the probed region. It can be seen that with increasing $h$-values, more and more plastic events occur inside the probed region $(r<5)$. As a reference, we also show the cumulative distribution for the original frozen matrix (FM) with $r=5$. The TM method with $h\to\infty$ reduces to FM as a special case, and we observe that already for $h=100$ the cumulative distributions of the plastic event distance $r$ are very close. A coupling strength $h=10$ still localizes about 60\% of all events inside the probe region, a value that drops to about 30\% for $h=0$. 

Panel (b) of Fig.~\ref{fig:distmod} shows the local shear moduli obtained from 10 independent systems comprised of 20,000 particles. They follow a normal distribution whose mean increases with increasing strength of the nonaffine suppression field $h$. While a value of $h=1$ does not appear to strongly perturb the measurement of the moduli, $h=100$ again yields a distribution quite similar to the frozen matrix method. These results suggests $1\le h\le 100$  as a reasonable range of values to use for our exploration of the local yield stress.  They are consistent with those of Mizuno et al. \cite{mizuno2013measuring}, who also found normal distributions of shear moduli. As in our calculations, the mean value shifts to larger values when a frozen constraint is employed. 

\begin{figure}[t]
\includegraphics[width=0.8\columnwidth]{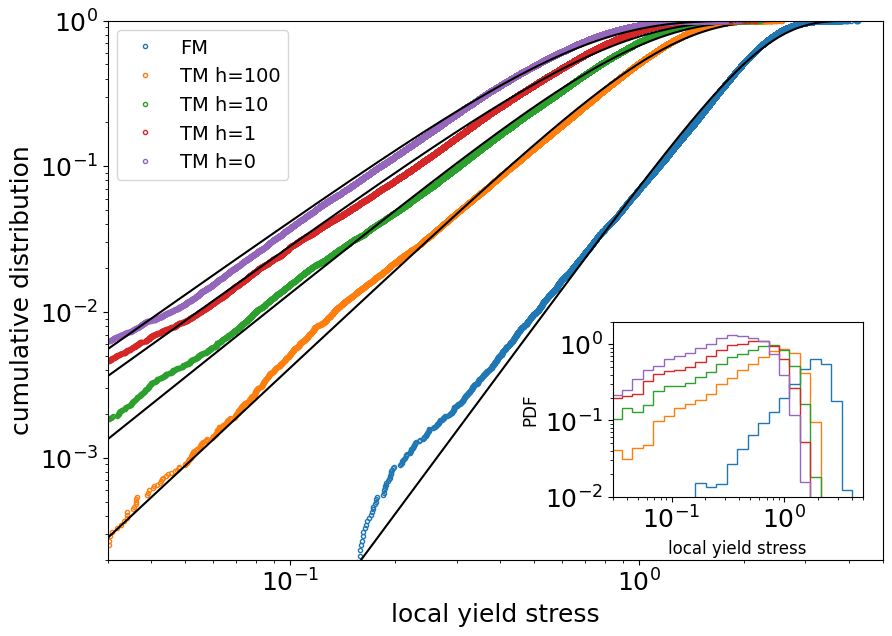}
\caption{\label{fig:lysdist} Cumulative probability distribution of local yield stresses $\Delta \sigma_y$. Black lines show fits to a Weibull distribution $C(\Delta \tau_y)=1-\exp[(\Delta \tau_y/\langle \Delta \tau_y \rangle)^{1+\theta}]$, with $\theta_{FM}=2.23$, $\theta_{h=100}=1.23$, $\theta_{h=10}=0.91$, $\theta_{h=1}=0.72$, $\theta_{h=0}=0.69$. The inset shows the probability distribution functions for the same data.}
\end{figure}

Having characterized the local linear properties of the material, we proceed by analyzing the distributions of local yield stresses of the same systems and for the same parameters as in Fig.~\ref{fig:distmod}. The main panel in Fig.~\ref{fig:lysdist} focuses on the cumulative distribution, which is better suited for studying the behavior at small values. We find that the probability distributions follow $P(\Delta \tau_y)\sim \Delta \tau_y^\theta$ for small yield stresses, i.e.~they vanish as a power law as expected for marginally stable solids. All cumulative distributions can be fitted with a Weibull form \cite{karmakar2010statistical}, $C(\Delta \tau_y)=1-\exp[(\Delta \tau_y/\langle \Delta \tau_y \rangle)^{1+\theta}]$. While this form is expected when extreme value statistics applies, it is not obvious that it should also hold for the characteristic yield stress of a small region. Relaxing the frozen matrix constraints decreases both the "pseudogap exponent" $\theta$ as well as the mean local yield stress. Similar to the elastic properties, results for $h=1$ approach those for $h=0$. With decreasing values of $h$, yields occur at smaller strains as the plastic events become less localized. 


The values predicted by the thawed matrix for the pseudogap expoent $\theta$ are substantially smaller than those from the frozen matrix, and asymptote to $\theta \approx 0.69$ as $h\rightarrow 0$. An independent and  nonpertubative way of estimating $\theta$ consists in measuring the dependence of the average strain $\langle \epsilon_y\rangle$ to the first failure event on the number of particles $N$ in the simulation, and invoke a scaling relation from extreme value statistics, $\langle \epsilon_y\rangle \sim N^{-1/(1+\theta)}$. Using this approach, prior work on very similar quenched model glasses reported $\theta=0.6$ (2D) \cite{karmakar2010statistical,hentschel2015stochastic} and $\theta=0.5$ (3D) \cite{shang2020elastic}. We did not repeat such a finite size scaling calculation for our specific systems here, but the true value of $\theta$ should be in the same range and definitely less than one. The thawed matrix estimates are therefore more accurate than those from the frozen matrix. The asymptotic value $\theta \approx 0.69$ approaches the estimates from earlier work, but the larger values of $\theta>1$ that we obtain for larger $h$ suggest that there are still residual artefacts.

\begin{figure}[t]
\includegraphics[width=\columnwidth]{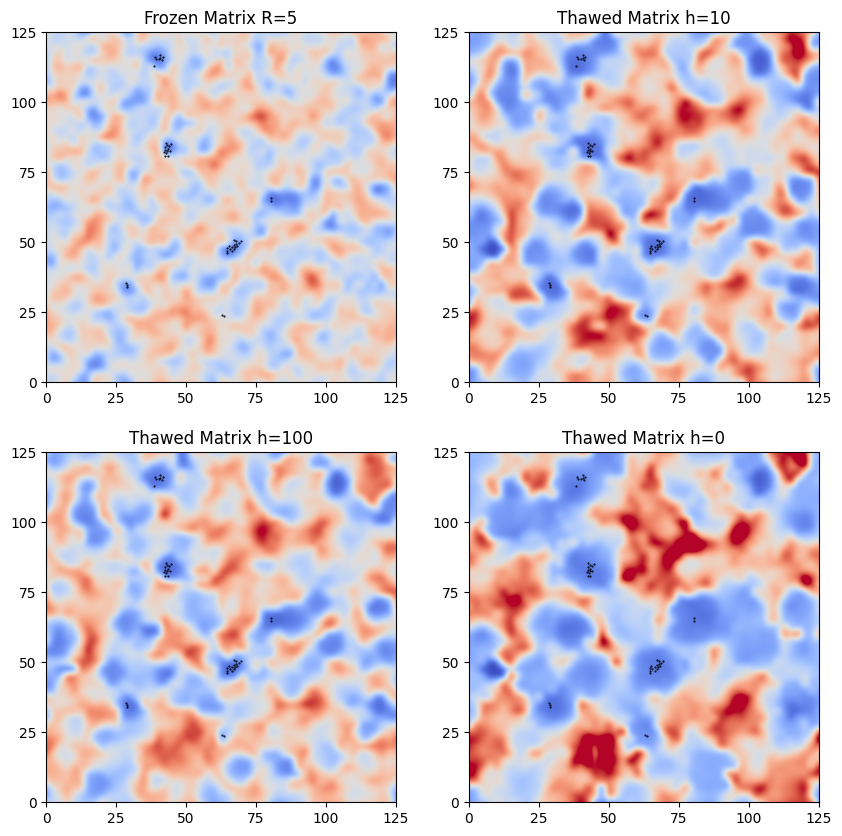}
\caption{\label{fig:map} Example map of local yield stresses, normalized by their mean values, in a system containing 20,000 particles. The colors range from 0 (blue) to 2 (red).  Black dots indicate particles with large plastic activity after 1\% strain, specifically those with an accumulated value of $D_{min}^2>0.5$.}
\end{figure}

In order to further characterize the effect of gradually "unfreezing" the environment around the probed region, we show in Fig.~\ref{fig:map} the local yield stress map of an example system, computed with the frozen matrix method as well as the thawed matrix with three different strengths of the nonaffine constraint field. It can be appreciated that the local yield stress exhibits substantial spatial correlations that cannot be seen in the distribution functions of Fig.~\ref{fig:lysdist}. Regions identified as weak (blue) or strong (red) by the frozen matrix tend to be labelled as such also in the thawed matrix, but their size grows with decreasing strength of the constraint field. This increase of the feature size with decreasing constraint field can be understood from the fact that with smaller values of $h$, the effective size of the local region being probed increases (see Fig.~\ref{fig:distmod}(a)).

Figure~\ref{fig:map} also shows as black dots the position of those particles that have experienced a plastic rearrangement (as defined by exceeding a threshold value of $D_{min}^2>0.5$)  when these systems are sheared globally (i.e. without any matrix constraints) up to a strain of 1\%. As in previous studies of the local yield stress \cite{patinet2016connecting}, these first few plastic events localize very precisely in the low local yield stress regions of all yield stress maps.

The important question we wish to answer now is whether our new thawed matrix method performs better in the prediction of plastic activity than the frozen matrix approach. To make this analysis specific, we turn to two observables that have previously been used in the literature to quantify such correlations \cite{patinet2016connecting,barbot2018local}. This allows for easy comparison with existing results. Of particular interest is to quantify how a map measured in an undeformed sample predicts plastic activity after a global shear strain $\gamma$. The first quantity is a correlation function defined via the cumulative distribution function (CDF) of the local yield stress,
\begin{equation}
    C(\gamma)=1-2\,CDF(\Delta \tau_{y,i}),
    \label{corr-eq}
\end{equation}
where $i$ denotes those particles that are considered plastically active in a strain interval around the strain $\gamma$. Since we have evaluated $\Delta \tau_y$ on a grid and not for every particle, we bin particles onto the same grid and accumulate their $D_{min}^2$ values over strain intervals $\Delta \gamma=0.05$. This yields a map of plastic activity for a given global strain $\gamma$ at the same spatial scale as the yield stress map. For the purpose of computing the correlation, a bin is considered plastic if its $D_{min}^2$ value exceeds a threshold of 0.5. Our procedure differs in some technical aspects from refs.~\onlinecite{patinet2016connecting,barbot2018local}, who used the maximum local shear strain $\sqrt{((\epsilon_{xx}-\epsilon_{yy})/2)^2+\epsilon_{xy}^2}$ as plastic indicator. We have verified, however, that our results are robust against this variation. Our second indicator is the Spearman rank correlation between maps of local yield stress and plastic activity at different values of $\gamma$. This metric obviates the need of a threshold to define plastic events.

\begin{figure}[t]
\includegraphics[width=\columnwidth]{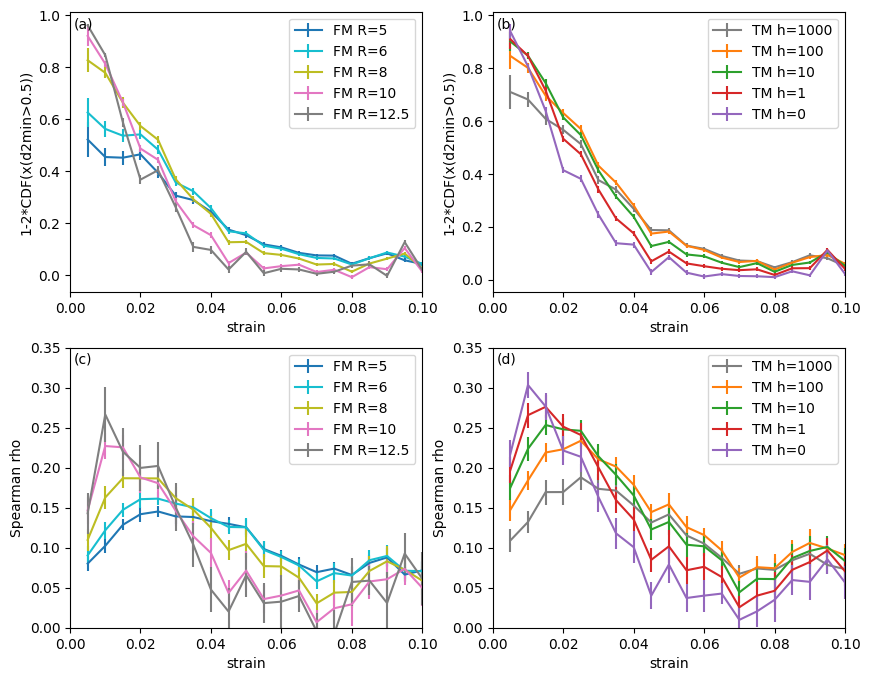}
\caption{\label{fig:corr} Correlation between local yield stress and plastic activity. Panels (a) and (b) show results for the correlation function eq.~\eqref{corr-eq}, while panels (c) and (d) show the Spearman rank coefficent. Left column: frozen matrix results, right column: thawed matrix results. Curves represent an average over 20 systems containing 20,000 particles. }
\end{figure}

Results for both of these correlation measures are reported in Fig.~\ref{fig:corr}, where data for the frozen matrix method (left column) for various sizes of the frozen region can be directly compared with data for the thawed matrix (right column) for different values of the coupling parameter $h$. The correlation function eq.~\eqref{corr-eq} decreases monotonically over a strain interval of 10\%, reflecting the progressive erasure of memory as the system passes from the transient loading period into steady state flow. The Spearman coefficient first rises to reach a maximum at a strain of a few percent, before decreasing to zero. The low correlation for small strains is due to the small number of plastic events in the elastic regime, while the decorrelation at large strains is due to the renewal of the yield stress map from plastic activity.

For the frozen matrix, our results reproduce very closely those of refs.~\onlinecite{patinet2016connecting,barbot2018local}: increasing the radius $r$ from 5 to 10 raises the correlation to nearly perfect for the first few plastic events, but also leads to a faster decay at larger strains. A larger frozen matrix does very well in finding the weakest site, but the entire yield stress map loses predictive power due to the loss of spatial resolution. The Spearman correlation coefficient reveals the same picture: as $r$ increases, the maximum correlation first increases and shifts to smaller strains, but then the map decorrelates more quickly.  

For the thawed matrix, all values of $h$ explored here yield a map that better predicts the first few plastic events (panel (b)). We also see that the map obtained with $h=100$ yields the slowest decorrelation from the plastic activity. The Spearman coefficients have the same general trend as for the frozen matrix, but are generally larger (panel (d)).

\begin{figure}[t]
\includegraphics[width=0.8\columnwidth]{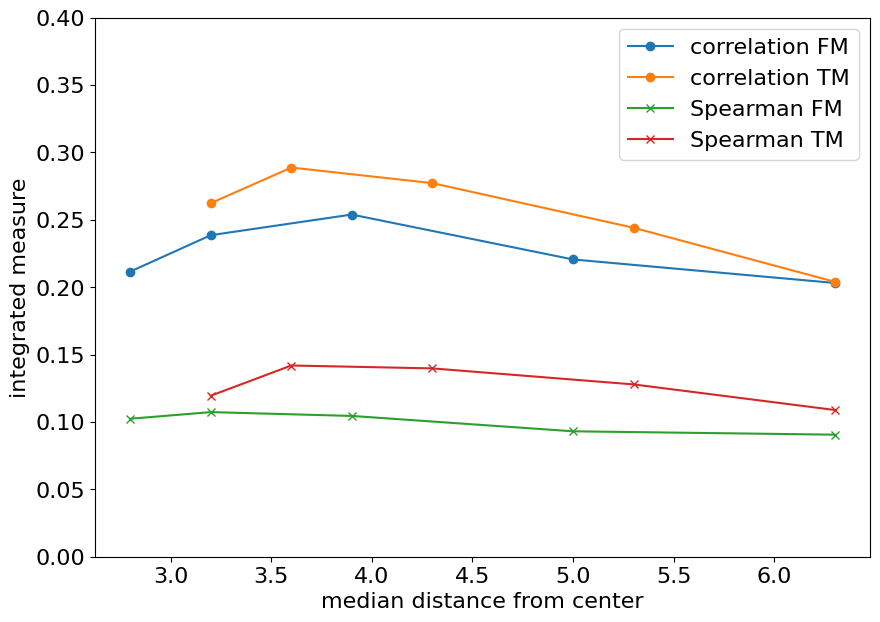}
\caption{\label{fig:intcorr}  Integrated correlation eq.~\eqref{intcorr-eq} between local yield stress and plastic activity from the data of Fig.~\ref{fig:corr}.}
\end{figure}

From these results, one can conclude that an accurate prediction of the initial plastic yielding is possible with both the thawed matrix for all $h$-values, and the frozen matrix when a value of $r>8$ is used. If one is only interested in this prediction, there is little advantage in using the thawed matrix approach. However, one may also be interested in how well the yield stress map predicts plastic activity further into the future. As a figure of merit that captures this aspect, Barbot et al.~\cite{barbot2018local} proposed to consider the integrated correlation function
\begin{equation}
    I=\frac{1}{\gamma^*}\int_0^{\gamma^*} C(\gamma )d\gamma,
    \label{intcorr-eq}
\end{equation}
which is a more global measure of the correlation over the entire range of strains. Results for this quantity (using $\gamma^*=0.1$) are shown in Fig.~\ref{fig:intcorr}. In order to compare results for frozen and thawed matrix methods on the same axis, we return to Fig.~\ref{fig:distmod}(b) and, for each value of radius $r$ or field $h$, define a 'localization length scale' from the median of the cumulative distributions. This median distance of the center of the probed region is used as the horizontal axis. It can be appreciated that the thawed matrix outpeforms the frozen matrix both for the correlation function and the Spearman coefficent, with $h=100$ yielding the best results.  We note that our results suggest an optimal size $r=8$ for the frozen matrix method, while  Barbot et al.~\cite{barbot2018local} found a smaller value $r=5$. These differences are likely due to the technical differences in the protocols for evaluating eq.~\eqref{corr-eq}.

\section{Conclusions}
We have introduced a new computational approach for the assessment of local plastic properties in amorphous solids. Our 'thawed matrix' approach no longer enforces strictly affine deformation as in the 'frozen matrix' method used to date, but localizes plastic activity through an external field that adds an energy cost for large local nonaffine deformations in the surrounding matrix. The strength of that field can be tuned to achieve a compromise between localization of plastic activity and artificially constraining relaxation. This 'thawed matrix' approach does not completely eliminate all artefacts, such as an overestimate of the pseudogap exponent, but leads to plasticity maps of quenched configurations that correlate better with plastic events occurring in the same systems under shear than the existing frozen matrix method.

From a conceptual point of view, we feel that the 'thawed matrix' can be seen as a computationally tractable approximation to an ``ideal'' measurement of local yield properties. We propose that the latter would consist of detecting the first plastic yield event in the region of interest, for a sample that is being sheared quasi-statically while continually minimizing its energy {\em subject to the maximum value of $D^2_{min}$ in the matrix not exceeding some fixed threshold}. With a suitably small threshold value, this would quantitatively implement the idea of suppressing all localised plastic rearrangements in the matrix while still allowing large-scale nonaffine displacements. Our thawed matrix method introduces a simplification by taking as the matrix not the entire system outside the region of interest, but a ring of finite size. This approximation could easily be removed in principle, though at the expense of higher computational cost. The $h$-field of the thawed matrix approach can be seen as a Lagrange multiplier implementing a constraint on the average value of $D^2_{min}$ in the matrix. A constraint on the maximum rather than the average could be approximated by summing higher powers of $D^2_{min}(k)$ in eq.~(\ref{tm-eq}), and this would be interesting to explore in future work. 

Given the recent 'soft matrix' proposal as an alternative method to address the same issue \cite{adhikari2023soft}, it would also be interesting to compare the performance of 'soft', 'thawed' and 'frozen' approaches quantitatively using the same descriptors and correlation functions. From a computational perspective, both 'soft' and 'thawed matrix' introduce an overhead from the application of the constraint forces. While those generated by eq.~\eqref{tm-eq} are more expensive to evaluate than a simple harmonic tethering to an affine background, one could hope that they are less blunt and therefore better able to strike the balance between suppression of plastic activity and allowing large-scale nonaffine relaxation in the matrix. We hope that our improved yield stress maps can make a useful contribution to the ongoing research effort linking structure and dynamics in glasses.

\begin{acknowledgments}
This research was supported in part through computational resources and services provided by Advanced Research Computing at the University of British Columbia. P.S. would like to dedicate this paper to the memory of Surajit Sengupta, whose original idea of using the non-affine field $h$ to constrain plastic activity was a key source of inspiration for this work.
\end{acknowledgments}

\section*{Data Availability Statement}
The data that support the findings of this study are available from the corresponding author upon reasonable request.

\nocite{*}
\bibliography{aipsamp}

\end{document}